\documentclass[11pt,a4paper]{article}

\usepackage{amsfonts,dsfont}
\usepackage[T1]{fontenc}
\usepackage{cite}
\usepackage{bbm}
\usepackage{indentfirst}
\usepackage{amsopn}
\usepackage{amsmath}
\usepackage{amssymb}
\usepackage{amsthm}
\usepackage{authblk}
\usepackage{hyperref}
\usepackage{graphicx}
\usepackage{color}
\usepackage{listings}
\usepackage{mathtools}
\usepackage{subfig}
\usepackage{enumerate}
\usepackage{footnote}

\usepackage{jlcode}
\definecolor{jlbase}{HTML}{000000}            % julia's base color
\definecolor{jlkeyword}{HTML}{000000}         % julia's keywords
\definecolor{jlliteral}{HTML}{000000}         % julia's literals
\definecolor{jlbuiltin}{HTML}{000000}         % julia's built-ins
\definecolor{jlcomment}{HTML}{000000}         % julia's comments
\definecolor{jlstring}{HTML}{000000}          % julia's strings
\definecolor{jlbackground}{HTML}{EEEEEE}      % the background of the code 
\definecolor{jlrule}{HTML}{000000}            % the rule of the code block

\makeatletter
\lstdefinestyle{mystyle}{
	basicstyle=%
	\ttfamily
	\lst@ifdisplaystyle\scriptsize\fi
}
\makeatother

\DeclareMathOperator{\tr}{tr}
\DeclareMathOperator{\Tr}{Tr}

\newcommand{\R}{\ensuremath{\mathbb{R}}}
\newcommand{\N}{\ensuremath{\mathbb{N}}}

\newcommand*{\suchthatOld}{%
	\;%
	\ifnum\currentgrouptype=16\relax% we are in a \left...\right
	\middle|%
	\else%
	\iftoggle{WithinBracMacro}{% we are in a \@Brac macro
		|% This works (change to some other char to see)
	}{% can't use \middle
		|%
	}% 
	\fi%
	\;%
}%

\newcommand{\ket}[1]{\ensuremath{\left|#1\right\rangle}}
\newcommand{\bra}[1]{\ensuremath{\left\langle#1\right|}}
\newcommand{\ketbra}[2]{\ensuremath{\ket{#1} \! \! \bra{#2}}}
\newcommand{\proj}[1]{\ensuremath{\ketbra{#1}{#1}}}
\newcommand{\braket}[2]{\ensuremath{\left\langle{#1}\suchthatOld{#2}\right\rangle}}

\newcommand{\1}{{\rm 1\hspace{-0.9mm}l}}

\newcommand{\SPAN}{\mathrm{span}}
\newcommand{\Lrm}{\ensuremath{\mathrm{L}}}
\newcommand{\Crm}{\ensuremath{\mathrm{C}}}

\newcommand{\CC}{\mathcal{C}}

\newcommand{\VV}{\mathcal{V}}
\newcommand{\LL}{\mathrm{L}}

\newcommand{\XX}{\mathcal{X}}
\newcommand{\s}{\mathcal{S}}

\newcommand{\YY}{\mathcal{Y}}

\newcommand{\ZZ}{\mathcal{Z}}

\newtheorem{theorem}{Theorem}
\newtheorem{fact}{Fact}

\newcommand{\appendixnumberline}[1]{Appendix\space}

\let\oldappendix\appendix
\makeatletter
\renewcommand{\appendix}{%
	\addtocontents{toc}{\let\protect\numberline\protect\appendixnumberline}%
	\renewcommand{\@seccntformat}[1]{Appendix~\csname the##1\endcsname:\quad}%
	\oldappendix
}
\makeatother

\begin{document}
	\title{Optimal representation of quantum channels }
	
	\author[1]{Paulina Lewandowska\footnote{plewandowska@iitis.pl}}
	\author[1]{Ryszard Kukulski}
	\author[1]{\L ukasz Pawela}
	\affil[1]{Institute of Theoretical and Applied Informatics, Polish 
		Academy of Sciences, ul. Ba{\l}tycka 5, 44-100 Gliwice, Poland}
	\date{}
	
	\maketitle
	\begin{abstract}
		This work shows an approach to reduce the dimensionality of matrix
		representations of quantum channels. It is achieved by finding a base 
		of the
		cone of positive semidefinite matrices which represent quantum 
		channels. Next,
		this is implemented in the \texttt{Julia} programming language as a 
		part of the
		\texttt{QuantumInformation.jl} package.
	\end{abstract}

	\section{Introduction} 
	This paper aims at finding an optimal representation of quantum channels for
	the purposes of machine learning. By \emph{optimal} we understand the lowest
	possible number of real parameters needed to define a quantum channel~\cite{holbrook2003noiseless}.
	Further, we would like this representation to be technically usable so that
	we could train, for instance, neural networks to approximate functions of
	this objects. Such a neural network could be used for computing the diamond
	norm~\cite{watrous2018theory} which in turn gives us a measure of  distance
	between quantum channels. This approach could provide a large speed boost in the problem of quantum channel discrimination~\cite{helstrom1976quantum,jenvcova2014base}. This problem is currently at the forefront of study in the field of quantum information theory.

	Our work is naturally divided into three parts. In the first part we show
	the mathematical structures needed to find the optimal representation. This
	involves dealing with cones of positive semidefinite matrices. The second
	part we present the example of whereas the last part presents the
	implementation of this example in the \texttt{Julia} language. This
	implementation is now a part of the
	\texttt{QuantumInformation.jl}~\cite{qizenodo,Gawron2018} numerical library
	available on-line at \href{https://
	github.com/iitis/QuantumInformation.jl}{\texttt{https://
	github.com/iitis/QuantumInformation.jl}}. Surprisingly, despite the complex
	mathematical structure and quite technical proofs, the implementation is
	relatively simple and therefore useful.

	\section{Mathematical framework}
	
	\subsection{Quantum channels}
	Let $\XX$, $\YY$ be complex finite-dimensional vector spaces, let $\Lrm(\XX,
	\YY)$ be the set of all linear operators transforming vectors from $\XX$ to
	$\YY$ and denote $\Lrm(\XX)\coloneqq\Lrm(\XX, \XX)$. Further, consider 
	mappings of the form
	\begin{equation}
	\Phi: \Lrm(\XX) \to \Lrm(\YY).
	\end{equation}
	The set of all such mappings will be denoted $\mathrm{T}(\XX, \YY)$ and 
	$\mathrm{T}(\XX)\coloneqq\mathrm{T}(\XX,
	\XX)$. Quantum channels are such $\Phi \in \mathrm{T}(\XX, \YY)$ which are 
	trace
	preserving and completely positive. The former means that
	\begin{equation}
	\forall A \in \Lrm(\XX) \;\;\; \Tr(\Phi(A)) = \Tr(A).
	\end{equation}
	The latter is a bit more complicated. Formally this condition can be written
	as
	\begin{equation}
	\forall \ZZ \;\; \forall A \in \Lrm(\XX \otimes \ZZ) \;\; A \geq 0 \implies 
	\left( \Phi \otimes \1_{\Lrm(\ZZ)} \right)(A) \geq 0.
	\end{equation}
	The intuitive explanation is as follows. First, consider a $\rho \in 
	\Lrm(\XX)$
	such that $\Tr(\rho) = 1$ and $\rho \geq 0$. Such an operator is called a
	quantum state. We would like our channels not only to transform states into
	states, but also we would like the ability to perform a channel on only a 
	part
	of the system. In other words we would like the output of $\left( \Phi 
	\otimes
	\1_{\Lrm(\ZZ)} \right)(\rho)$ to also be a proper quantum state for an 
	arbitrary
	space $\ZZ$ and all $\rho \in \Lrm(\XX \otimes \ZZ)$. This can only be 
	fulfilled
	when we introduce the need for completely positivity. We will denote the 
	set of
	all quantum channels as $\Crm(\XX, \YY)$ and $\Crm(\XX) = \Crm(\XX, \XX)$.
	
	The mappings $\mathrm{T}(\XX, \YY)$ may be represented in a number of ways. 
	For our
	purposes only the Choi-Jamio{\l}kowski 
	isomorphism~\cite{choi1975completely,jamiolkowski1972linear} will be relevant. This representation 
	states that there exists a bijection
	$J$ between the sets $\mathrm{T}(\XX, \YY)$ and $\Lrm(\YY \otimes \XX)$. 
	This bijection
	can be explicitly written as
	\begin{equation}
	J(\Phi) = \sum_{i, j}^{\dim(\XX)} \Phi(\ketbra{i}{j}) \otimes \ketbra{i}{j}.
	\end{equation}
	$\Phi$ is completely positive if and only if $J(\Phi)\geq 0$; $\Phi$ is 
	trace
	preserving if and only if $\Tr_\YY J(\Phi) = \1_\XX$. Finally, $\Phi$ is
	Hermiticity preserving if and only if $J(\Phi) \in \mathrm{Herm}(\YY \otimes
	\XX)$, where $\mathrm{Herm}(\XX)$ denotes the set of all Hermitian matrices 
	in
	$\Lrm(\XX)$.
	
	\subsection{Convex cone structures}
	Consider $\XX$ is a real finite-dimensional vector space and $\CC \subset 
	\XX$ is
	a  closed  convex  cone. We  assume  that $\CC$ is pointed,  i.e. $\CC \cap 
	-\CC
	= \{ 0 \}$ and generating,  i.e. for each $ x \in \XX$ there exists $ u, w 
	\in
	\CC$ such that $ x = u - w$. Such a cone $\CC$ is called a proper cone in 
	the
	space $\XX$. The proper cone $\CC$ becomes a partially ordered vector space 
	$ x
	\ge y \iff x-y  \in \CC$ for each $x,y \in \XX$. Let $\XX^*$ be the space 
	dual to $\XX$ defined by the inner product $\braket{\cdot}{\cdot}$.  Then, 
	we may
	introduce a partial order in $\XX^*$ as well with the dual cone \begin{equation}
	\CC^* = 
	\{f
	\in \XX^*: \braket{f}{z} \ge 0, \forall z \in \CC \}.
	\end{equation}  The cone $\CC^*$ is 
	also
	closed and convex cone. If $\CC$ is generating in space $\XX$, then $\CC^*$ 
	is
	pointed and we may introduce partial order in $\XX^*$ given by
	\begin{equation}
	f \ge g \iff f-g \in \CC^*
	\end{equation}
	for all $f,g \in \XX^*$. 
	
	An interior point $e \in \mathrm{int}(\CC)$ of a cone $\CC$ is called an 
	order
	unit~\cite{fuchssteiner2011convex} if for each $ x \in \XX$, there exists  $ \lambda >0 $ such that $
	\lambda e - x \in \CC$ whereas a base of $\CC$ is defined as compact and convex
	subset $B \subset \CC$ such that for every $z \in \CC\setminus\{0\}$,
	there exists unique $t > 0$  and an element $b \in B$ such that $ z =
	tb. $ The following theorem shows there exists relation between the order
	unit $e$ and a base of cone $\CC$.
	
	\begin{theorem}\label{baza}
		The set  $B_e = \{z \in \CC: \braket{e}{z} = 1 \} $ is the base of $\CC$
		(determined by element $e$) if and only if an element $e$  is an order 
		unit and $e\in \mathrm{int}\left(\CC^*\right)$.
	\end{theorem}
	The proof of this theorem is presented in Appendix~\ref{app:a}.

	\subsection{Base of Hermiticity preserving maps}
	Let us now define the finite-dimensional linear space 
	\begin{equation}\label{space}
	\{\Phi \in \mathrm{T}(\XX,\YY): \Phi \text{ -- Hermiticity preserving} \}.
	\end{equation}
	Due to the Choi--Jamiolkowski isomorphism, the set of all Hermiticity 
	preserving
	linear maps of a finite-dimensional space is  mathematically  closely 
	related to the~set 
	\begin{equation}
	\VV = \{J(\Phi): J(\Phi) \in \mathrm{Herm}(\YY \otimes \XX) \},
	\end{equation}	 of all Choi matrices of Hermiticity preserving maps.

	In every linear space of Hermitian matrices $\mathrm{Herm}(\ZZ)$ we can 
	introduce an orthonormal basis $\mathcal{B}(\ZZ)$. The basis 
	$\mathcal{B}(\ZZ)$ is a collection of $\dim(\ZZ)^2$ matrices. The standard 
	orthonormal basis is denoted by the set
	\begin{equation}
	\begin{split}
	\mathcal{B}(\ZZ) = & \\ 
	& \left\{
	\frac{\1_\ZZ}{\sqrt{\dim(\ZZ)}}, \right .\\
	& \frac{\sum_{a=1}^k \proj{a} - k 
		\proj{k+1}}{\sqrt{k+k^2}}, \mbox{ for } k=1,\ldots,\dim(\ZZ)-1, \\
	& \left. \frac{\ketbra{a}{b}+\ketbra{b}{a}}{\sqrt{2}}, \frac{i 
	\ketbra{a}{b}-i
		\ketbra{b}{a}}{\sqrt{2}}, \mbox{ for } a,b=1,\ldots,\dim(\ZZ) \mbox{ 
		and } a\not=b \right\}.
	\end{split}
	\end{equation}
	If we consider the space $\VV$ of all Choi matrices of Hermiticity 
	preserving maps we receive the $\dim(\XX)^2 \dim(\YY)^2 $ dimensional 
	space. To reduce the number of dimensions of $\VV $   we introduce the 
	concept of a cone in this space and the  base of cone.

	Now we introduce a proper cone in the space $\VV$ as
	\begin{equation}
	\CC = \{ J(\Phi) \in \VV: J(\Phi) \ge 0 \},
	\end{equation}

	and  a subspace $\s \subset \VV$ such that
	\begin{equation}
	\s  = \{J(\Phi) \in \VV : \tr_{\YY} J(\Phi) = c \1_\XX \, , \, c \in \R \}.
	\end{equation}
	By $\s^{\perp}$ we denote the orthogonal complement of $\s$ which is given 
	by
	\begin{equation}
	\s^{\perp} \coloneqq \{ X \in \VV: \tr \left( X Y \right)  = 0, Y \in \s  
	\}.
	\end{equation}
	\begin{fact}\label{fact1}
		The set $\s^{\perp}$ is given by
		\begin{equation}
		\s^{\perp} = \{ \1_\YY \otimes H :H \in \mathrm{Herm}(\XX), \tr(H) = 0 
		\}.
		\end{equation}
	\end{fact}
	
	The proof of this fact is presented in Appendix~\ref{app:b}.
	
	We can also introduce a proper cone $\CC_\s$ in space $\s$ given by $\CC_\s 
	= \s
	\cap \CC$ and a base $B_\s \subset \CC_\s$ of the cone $\CC_\s$. We can 
	prove,
	using Theorem~\ref{baza}, that the set $B_\s$ is the base of cone $\CC_\s$ 
	if
	and only if  $B_\s = \s \cap B_E$ for some order unit $E \in
	\mathrm{int}(\CC^*)$. The base $B_\s$  determined by an order unit $E $ 
	will be
	denoted as $ B_\s^E $ and is given by
	\begin{equation}
	B_\s^E= \{ X \in \CC_\s: \braket{X}{E}=1 \}.
	\end{equation}
	One can easily see that identity matrix $\1_\YY \otimes \1_\XX$ is an order 
	unit in cone $\CC$. Thus we have the following observation.

	\begin{fact}\label{fact2}
		For $E\coloneqq \frac{\1_\YY \otimes \1_\XX}{\dim(\XX)}$ the base 
		$B^E_\s$ is determined by the set of Choi matrices of quantum channels 
		$\Phi \in \mathrm{C}(\XX,\YY)$ i.e.
		\begin{equation}
		B^E_\s = \{ J(\Phi): \Phi \in \mathrm{C}(\XX,\YY) \}.
		\end{equation}
	\end{fact}
	
	We are ready to establish the main result of our work.
	
	\begin{theorem}\label{th-basis}
		The linear space $\s$ is the smallest linear subspace containing the 
		set of 
		quantum channels $\mathrm{C}(\XX,\YY)$ which orthonormal basis 
		$\mathcal{B}(\s)$ given by
		\begin{equation}
		\left\{\frac{\1_\YY \otimes \1_\XX}{\sqrt{\dim(\XX) 
				\dim(\YY)}}\right\} \cup \left\{G \otimes H: G \in 
		\mathcal{B}(\YY)\backslash \left\{\frac{\1_\YY}{\sqrt{\dim(\YY)}} 
		\right\} 
		, H \in 
		\mathcal{B}(\XX)\right\}.
		\end{equation}
		Moreover, 
		\begin{equation}
\dim(\s)=\dim(\XX)^2 \dim(\YY)^2 - \dim(\XX)^2 + 1.
		\end{equation}
	\end{theorem}
The proof of this theorem is presented in Appendix~\ref{app:c}.

	Theorem \ref{th-basis} states that every quantum channel $\Phi \in 
	\mathrm{C}(\XX,\YY)$ can be uniquely determined by $\dim(\XX)^2 \dim(\YY)^2 
	- 
	\dim(\XX)^2$ real numbers due to fact that for every $\Phi \in 
	\mathrm{C}(\XX,\YY)$ holds $\tr_{\YY} J(\Phi) = \1_\XX$. Moreover, for 
	$J(\Phi)$ the coefficient
	\begin{equation}
	\braket{\frac{\1_\YY \otimes \1_\XX}{\sqrt{\dim(\XX) 
				\dim(\YY)}}}{J(\Phi)}=\sqrt{\frac{\dim(\XX)}{\dim(\YY)}}.
	\end{equation}
	remains fixed for every $\Phi \in \mathrm{C}(\XX, \YY)$.
	
	As a conclusion, we reduced the dimension of computational space by 
	$\dim(\XX)^2$. 
	
	\section{Example}
	
	In this section we present how one can use the \texttt{Julia} language and \linebreak
	\texttt{QuantumInformation.jl} library in order express quantum channels as 
	vectors in the space $S$.
	
	Let us consider $\XX=\mathbb{C}^2$ and $\YY=\mathbb{C}^3$ along with quantum
	channels 
	$\Phi \in \mathrm{C}(\XX)$ given by
	\begin{equation}\label{eq}
	\begin{split}
	\Phi(X)&=\frac{1}{2} \left[\begin{matrix}
	1 & 1\\
	1 & -1
	\end{matrix}\right] X \left[\begin{matrix}
	1 & 1\\
	1 & -1
	\end{matrix}\right], \quad X \in \LL(\XX),
	\end{split}
	\end{equation}
	and  $\Psi \in \mathrm{C}(\YY)$ defined as
	\begin{equation}
	\begin{split}
	\Psi(Y)&=\left[\begin{matrix}
	1 & 0.92-0.14i  & 0.84-0.19i\\
	0.92+0.14i&  1 & 0.81+0.06i\\
	0.84+0.19i & 0.81-0.06i & 1\\
	\end{matrix}\right] \odot Y, \quad Y \in \LL(\YY),
	\end{split}
	\end{equation}
	where $\odot$ denotes the Hadamard product.
	
	First we calculate the Choi matrices of $\Phi$ given by
	\begin{equation}
	J(\Phi)=
	\left[\begin{matrix}
	0.5 &  0.5 &  0.5 & -0.5 \\
	0.5 &  0.5 &  0.5 & -0.5 \\
	0.5 &  0.5 &  0.5 & -0.5 \\
	-0.5 & -0.5 & -0.5 &  0.5
	\end{matrix}\right].\end{equation} Analogously for  $\Psi$  we have \begin{equation}
	J(\Psi)= 
	\left[\begin{matrix}
	1   &    0 & 0 & 0 & 0.92-0.14i& 0  &0&  0 & 0.84-0.19i \\
	0    &   0 & 0 & 0  &     0  &      0 & 0 & 0   &   0       \\
	0  &    0  &0&  0  &     0 &       0&  0&  0  &     0  \\
	0  &     0 & 0 & 0  &     0   &     0 & 0 & 0  &     0 \\
	0.92+0.14i& 0&  0 & 0  &     1 &       0 & 0  &0&  0.81+0.06i\\
	0    &   0&  0  &0   &    0   &     0 & 0&  0  &     0 \\
	0    &   0  &0&  0   &    0   &   0  &0 & 0  &    0 \\
	0   &    0 & 0 & 0  &    0   &     0  &0 & 0   &    0 \\
	0.84+0.19i & 0  &0 & 0 &0.81-0.06i  &0 & 0 & 0   &    1 \\    
	\end{matrix}\right].
	\end{equation}
	Now we use the function \texttt{channelbasis}. The inputs of this 
	function are the dimensions of spaces $\XX$ and $\YY$ of channels $\Phi, 
	\Psi$. The function returns an orthonormal basis 
	of $S$. Then, we are able to use the function \texttt{represent} which 
	factor 
	out Choi matrices $J(\Phi), J(\Psi)$ on basis elements and returns a vector 
	representations 
	$v_{J(\Phi)}, v_{J(\Psi)}$  of basis coefficients. In our examples 
	we have
	
	\begin{equation}
	v_{J(\Phi)}=\left[\begin{matrix}
	1.0    \\
	0.70711\\
	0.70711\\
	-0.70711\\
	-0.70711\\
	1.0 
	\end{matrix}\right] \oplus \mathbf{0}_7, \quad v_{J(\Psi)}= 
	\left[\begin{matrix}
	0.70711 \\  -0.70711\\ -0.20356 \\-0.26978\\ 1.29553 \\0.40825 \\0.40825\\ 
	-0.8165 \\0.08119\\ 1.18231\\ 1.14206 \\1.0
	\end{matrix}\right] \oplus \mathbf{0}_{61}
	\end{equation}
	where $\mathbf{0}_i$ denotes vector of zeros of length $i$.
	If we want to reverse vector representation process, 
	we can use function 
	\texttt{combine}. The output matrix elements shall be accurate with 
	original 
	Choi matrix elements to $10^{-16}$ or better.
	\section{Julia implementation}
	Here, we present the code structure for the basis 
	representation of Choi matrix of a qubit unitary channel $\Phi$ given by 
	Eq. 
	\eqref{eq}.
	
	\begin{lstlisting}
	julia> using QuantumInformation
	
	julia> H=hadamard(2)
	2×2 Array{Float64,2}:
	0.707107   0.707107
	0.707107  -0.707107
	
	julia> # defining Choi Matrix
	J_Φ=res(H)*res(H)'
	4×4 Array{Float64,2}:
	0.5   0.5   0.5  -0.5
	0.5   0.5   0.5  -0.5
	0.5   0.5   0.5  -0.5
	-0.5  -0.5  -0.5   0.5
	
	julia> # representing Choi matrix in the basis of the subspace S
	v_J_Φ=represent(channelbasis(Matrix{ComplexF64}, 2, 2),J_Φ)
	13-element Array{Float64,1}:
	0.0               
	0.9999999999999996
	0.0               
	0.0               
	0.0               
	0.0               
	0.0               
	0.0               
	0.7071067811865474
	0.7071067811865474
	-0.7071067811865474
	-0.7071067811865474
	0.9999999999999998
	
	julia> # recovering the original Choi matrix from its basis representation
	J_Φ_recovered=combine(channelbasis(Matrix{ComplexF64}, 2,2),v_J_Φ).matrix
	4×4 Array{Complex{Float64},2}:
	0.5+0.0im   0.5+0.0im   0.5+0.0im  -0.5+0.0im
	0.5+0.0im   0.5+0.0im   0.5+0.0im  -0.5+0.0im
	0.5+0.0im   0.5+0.0im   0.5+0.0im  -0.5+0.0im
	-0.5+0.0im  -0.5+0.0im  -0.5+0.0im   0.5+0.0im
	
	julia> # checking accuracy of recovery process using trace norm
	print(norm_trace(J_Φ-J_Φ_recovered))
	8.881784197001252e-16
	\end{lstlisting}

	\section{Conclusion}
	In this work we  find a matrix basis for quantum channels and provide  strict mathematical proofs supporting our result. This basis allows us to reduce the dimensionality of the matrix which represents a quantum channel. This, in turn, allows us to speed up computation of a class of functions of these channels, which is applicable in, for instance, the study of quantum channel discrimination. Our analytical results are accompanied by functions written in the~\texttt{Julia} language which decompose a given quantum channel in our basis. This implementation is now a part of the~\texttt{QuantumInformation.jl} package~\cite{qizenodo,Gawron2018}.
	
		\section*{Acknowledgements}
	
	This work was supported by the Foundation for Polish Science (FNP) under grant 
	number POIR.04.04.00-00-17C1/18-00.

	\appendix
	\section{Proof of Theorem~\ref{baza}}\label{app:a}
	\begin{proof}
		$"\Longrightarrow"$ Consider that $B_e$ is a base of $\CC$. An element 
		$e 
		\in \mathrm{int}(\CC^*)$ if and only if there exists $r>0$ such that 
		the 
		ball $ 
		K(e,r) \subset \CC^*$, which is equivalent above condition
		\begin{equation}
		\exists_{ r>0} \forall_{f \in \XX^* } \left( ||f-e|| < r \implies 
		\forall_{z \in \CC} \braket{f}{z} \ge 0  \right).
		\end{equation}
		By using the fact that in finite-dimensional spaces all norms are 
		equivalent, we use the definition of induced norm given by  
		\begin{equation}
		||f-e|| = \inf \left\{ M \in [0,\infty): \forall_{x \in \XX } 
		|\braket{f}{x} - \braket{e}{x}| \le M ||x|| \right\}.
		\end{equation}
		Then, we have
		\begin{equation}
		||f-e|| < r \iff \exists_{ 0<M<r} \forall_{x \in \XX} |\braket{f}{x} - 
		\braket{e}{x}| \le M ||x||. 
		\end{equation}
		Assume that $f \in~K(e,r)$, $M \coloneqq \max\{ ||b||: b \in B_e\}$ and 
		$0 < r < 
		\frac{1}{M}$. Then, we have
		\begin{equation}
		|\braket{f}{b} - \braket{e}{b}| \le ||f-e||\cdot||b|| \le  r||b|| \le 
		rM <1.
		\end{equation} 
		If $b\in B_e$, then $\braket{e}{b}=1$. 
		Hence $
		|\braket{f}{b}-1|<1. $ That entails that $\braket{f}{b} >0 $. By using 
		the 
		assumption we have $\braket{f}{z} = t \braket{f}{b}$, which implies 
		that 
		$\braket{f}{z} > 0$. 
		
		$"\Longleftarrow"$ Now consider that $e \in  \mathrm{int}(\CC^*)$ is 
		order
		unit. It easy to see that $B_e$ is a convex set. First prove that
		$\braket{e}{z} \neq 0$. Let $z \in \CC \setminus \{ 0 \}$ and $e \in
		\mathrm{int}(\CC^*)$. If $e \in \CC^*$, then $\braket{e}{z} \ge 0$. It
		suffices to show that $\braket{e}{z} \neq 0$. We will show this fact by
		contradiction. Assume $\braket{e}{z} = 0$ and let $\epsilon > 0 $.  By 
		the
		Hahn--Banach theorem \cite{rudin1964principles}, there exists $z^* \in
		\XX^*$ such that $\braket{z^*}{z} = ||z||$. Then
		\begin{equation}
		\braket{e-\epsilon z^*}{z} = -\epsilon ||z|| < 0 
		\end{equation}
		It implies that $K(e, \epsilon) \not\in \CC^*$, which is contradiction 
		with
		the assumption $e \in \mathrm{int}(\CC^*)$. Therefore, $\braket{e}{z} > 
		0$.

Let us see	that if	$b\coloneqq \frac{z}{\braket{e}{z}}$ and $t \coloneqq
		\braket{e}{z}$, then each element $z \in \CC \setminus \{ 0 \}$  can be 
		written
		as $z = tb$. To prove that $B_e$ is compact we note that $\XX$ is a
		finite-dimensional space. Then, the set $B_e$ is compact if and only if
		$B_e$ is closed and bounded. To prove that $B_e$ is closed, take any
		sequence $(z_n)_{n \in \N} \in B_e$ such that $z_n \xrightarrow{n
			\rightarrow \infty } z$. By the inner product continuity, we get
		\begin{equation}
		1 = \lim\limits_{n \rightarrow \infty} \braket{e}{z_n} = 
		\braket{e}{\lim\limits_{n \rightarrow \infty} z_n} = \braket{e}{z}.
		\end{equation}
		It implies that $z \in B_e$ therefore $B_e$ is closed. To prove that 
		$B_e$
		is bounded we show there exists $M \in [0,\infty)$ such that $||z|| \le 
		M$
		for every $z \in B_e$. Let us take a compact sphere $\mathrm{S}(0,1)$  
		and
		closed cone $\CC$. Then $ S = \mathrm{S}(0,1) \cap \CC$ is also compact.
		Notice the function $f: S \rightarrow \R_+ $ given by $f(x) = 
		\braket{e}{x}
		$, where $e$ is an order unit. By the Weierstrass theorem, a function 
		$f$
		attains infimum and supremum. Therefore, there exists $x_0 \in S$ such 
		that
		$ 0\le f(x_0) = \inf_{x \in S} f(x)$. Consider by contradiction that 
		$f(x_0)
		= \braket{e}{x_0} = 0$. We have $0 = \braket{e}{x_0} = \braket{e}{tb_0} 
		=
		t$, where $b_0 \in B_e$, which is a contradiction with the assumption 
		$t > 0
		$. Thus there exists $\lambda\coloneqq\braket{e}{x_0} > 0 $ such that $
		\braket{e}{z} \ge \lambda ||z||$  for every $z \in B_e$, hence $ ||z|| 
		\le
		\frac{1}{\lambda}.$ Taking $M\coloneqq \frac{1}{\lambda}$, we get 
		thesis.
	\end{proof}
	
	\section{Proof of Fact~\ref{fact1}}\label{app:b}
	
	\begin{proof}
		It is clear that $\dim(\VV) =(\dim(\XX) \dim(\YY))^2$. Consider a 
		linear 
		space $\VV \oplus \R$ which is $(\dim(\XX) \dim(\YY))^2+1$ dimensional. 
		Take any $J(\Phi) \in \s$. The condition  $\tr_{\YY} J(\Phi) = c 
		\1_\XX, c \in 
		\R $ in the space $\VV \oplus \R$  is equivalent to
		\begin{equation}
		\sum_{k=1}^{\dim(\YY)} \Re \left(J(\Phi)_{j+(k-1)\dim(\XX), 
			i+(k-1)\dim(\XX)} \right) - c = 0  \,\,\,\,\,\, \forall_{i,j \in 
			\{1,\ldots, \dim(\XX) \}} 
		\end{equation}
		for all $i,j \in \{1,\ldots, \dim(\XX) \}$. This homogeneous system of
		$\dim(\XX)^2$ linear equations is linearly independent. By rank--nullity
		theorem~\cite{meyer2000matrix}, we have
		\begin{equation}
		\dim(\s)=(\dim(\XX) \dim(\YY))^2+1 - \dim(\XX)^2.
		\end{equation}
		Therefore,  $\dim(\s^{\perp}) = \dim(\XX)^2-1$. To complete the proof, 
		note 
		that \begin{equation}
		\dim \left(\{ \1_\YY \otimes H :H \in 
		\mathrm{Herm}(\XX), 
		\tr(H) =0\} \right) =  \dim(\XX)^2-1.
		\end{equation}
	\end{proof}
	
	\section{Proof of Theorem~\ref{th-basis}}\label{app:c}
	
	\begin{proof}
		According to Fact \ref{fact2} the set $\mathrm{C}(\XX,\YY)$ is the base 
		of 
		a proper cone $\CC_\s$. That means
		\begin{equation}
		\SPAN\left( \{J(\Phi) \in \mathrm{Herm}(\YY \otimes \XX) \}  \right)=\s.
		\end{equation}
		Now we fix na orthonormal basis of the space $\VV$. Let it be given as 
		the collection
		\begin{equation}
		\mathcal{B}(\VV) = \{G \otimes H: G \in \mathcal{B}(\YY), H 
		\in 
		\mathcal{B}(\XX) \}.
		\end{equation}
		By using Fact~\ref{fact1}, if we take $X \in 
		S^\perp$, than there 
		exists $H \in \mathrm{Herm}(\XX)$, $\tr(H)=0$ such that $X=\1_\YY 
		\otimes 
		H$. Let us set up the basis of $S^\perp$
		\begin{equation}
		\mathcal{B}(\s^\perp)=\left\{\frac{\1_\YY}{\sqrt{\dim(\YY)}} \otimes 
		H: H \in 
		\mathcal{B}(\XX)\backslash \left\{ \frac{\1_\XX}{\sqrt{\dim(\XX)}} 
		\right\} 
		\right\}.
		\end{equation}
		Bearing in mind the relation $\VV=\s\oplus \s^\perp$, we conclude that 
		basis 
		of $\s$ can be chosen as $\mathcal{B}(\s)=\mathcal{B}(\VV) 
		\backslash \mathcal{B}(\s^\perp)$, namely
		\begin{equation}
		\left\{\frac{\1_\YY \otimes \1_\XX}{\sqrt{\dim(\XX) 
				\dim(\YY)}}\right\} \cup \left\{G \otimes H: G \in 
		\mathcal{B}(\YY)\backslash \left\{\frac{\1_\YY}{\sqrt{\dim(\YY)}} 
		\right\} 
		, H \in 
		\mathcal{B}(\XX)\right\},
		\end{equation}
		which completes the proof.
	\end{proof}
\end{document}